\documentclass[twocolumn,prl,showpacs,preprintnumbers,amssymb,superscriptaddress]{revtex4-1}

\usepackage{epsfig, color}
\usepackage{amsthm, amsmath, amsfonts, ae, psfrag}

\newcommand{\upd}{\mathrm{d}}
\newcommand{\bvec}[1]{{\bf\string#1}}

\begin{document}

\title{Dynamical density functional theory for dense suspensions of colloidal hard spheres}

\author{Daniel~Stopper}

\email{daniel.stopper@uni-tuebingen.de}

\author{Roland~Roth}

\author{Hendrik~Hansen-Goos}

\affiliation{Institute for Theoretical Physics, University of T\"ubingen, Auf der Morgenstelle 14, 72076 T\"ubingen, Germany
}

\date{\today}

\begin{abstract}

We study structural relaxation of colloidal hard spheres undergoing Brownian motion using dynamical density functional theory. Contrary to the partial linearization route [Stopper {\em et al.}, Phys. Rev. E {\bf 92}, 022151 (2015)] which amounts to using different free energy functionals for the self and distinct part of the van Hove function $G(r,t)$, we put forward a unified description employing a single functional for both components. To this end, interactions within the self part are removed via the zero-dimensional limit of the functional with a quenched self component. In addition, we make use of a theoretical result for the long-time mobility in hard-sphere suspensions, which we adapt to the inhomogeneous fluid. Our results for $G(r,t)$ are in excellent agreement with numerical simulations even in the dense liquid phase. In particular, our theory accurately yields the crossover from free diffusion at short times to the slower long-time diffusion in a crowded environment.

\end{abstract}

\maketitle


Structural properties of colloidal systems in equilibrium have been successfully described by classical density functional theory (DFT) \cite{Ev79,GrueKl99,ThorEA14}, which provides a powerful tool for the study of correlation functions in the presence of inhomogeneities caused by external potentials. In particular, colloids with hard-sphere like interactions are very accurately described by fundamental measure theory (FMT), a free energy model for the non-uniform hard sphere fluid including mixtures \cite{Rf89}. Attractive interactions can be modeled within FMT via depletion forces in a colloid-polymer mixture \cite{SchBrCP00}.
Recently, FMT has been employed within dynamical density functional theory (DDFT) \cite{MaTa99,ArEv04} in order to study the van Hove distribution function of colloidal suspensions undergoing Brownian motion \cite{Stopper15}, thereby extending the application of FMT into the realm of non-equilibrium systems. While the results were significantly improved compared to an approach using a cruder free energy model \cite{HopEA10}, two shortcomings were observed: (i) interactions within the self component of the van Hove function had to be removed using a partial linearization approach, effectively introducing different free energy functionals for each of the two components, and (ii) in the long time limit the theory yields the diffusion coefficient of the ideal gas, which is inconsistent with Brownian motion in a crowded environment. In this Communication, we solve both problems by (i) removing self interactions using a quenched self component in the 0D limit, which leads to a unified formulation that uses a single functional, and (ii) introducing a particle mobility which is a function of a locally averaged packing fraction, leading to a theory that yields the correct short-term {\em and} long-term diffusion coefficient.


We begin with a brief review of DDFT, which has been put forward originally by Marconi and Tarazona \cite{MaTa99} and Archer and Evans \cite{ArEv04}. We focus on the DDFT for a binary mixture featuring the self and distinct components with density profiles $\rho_s(\bvec{r},t)$ and $\rho_d(\bvec{r},t)$, denoted by $\rho_{s\!/\!d}(\bvec{r},t)$ in the following, which correspond directly to the respective parts of the van Hove distribution function $G_{s\!/\!d}(\bvec{r},t)$. The starting point for DDFT is the equilibrium free energy functional $\mathcal{F}[\rho_{s},\rho_{d}]$ of the mixture, conveniently of the FMT type in the case of the hard-sphere fluid. In analogy with the equilibrium chemical potentials $\mu_{s\!/\!d} = \frac{\partial f}{\partial \rho_{s\!/\!d}}$, where $f$ is the equilibrium free energy density of the bulk system, we can define local chemical potentials in a non-equilibrium configuration via
\begin{equation}
\label{eq_defddft1}
  \mu_{s\!/\!d}(\bvec{r},t) = \frac{\delta \mathcal{F}[\rho_{s},\rho_{d}]}{\delta \rho_{s\!/\!d}(\bvec{r},t)} \, .
\end{equation}
If we assume the particle current to be driven by chemical potential gradients, which is consistent with Brownian motion of non-interacting particles, we find the particle current densities
\begin{equation}
\label{eq_defddft2}
  \bvec{j}_{s\!/\!d}(\bvec{r},t) = -\Gamma_{s\!/\!d}(\bvec{r},t) \rho_{s\!/\!d}(\bvec{r},t) \nabla \mu_{s\!/\!d}(\bvec{r},t) \, ,
\end{equation}
where we have introduced the particle mobilities $\Gamma_{s\!/\!d}$. For non-interacting particles this is obviously a constant $\Gamma_0$ but in a system with non-trivial interactions $\Gamma_{s\!/\!d}$ should be allowed to depend on spatial coordinates and on time. Such a dependence has been implemented for the description of hydrodynamic interactions (see Ref.~\onlinecite{EsLoe09} and references therein). 

Finally, in order to obtain equations for the time evolution of the density profiles, it suffices to employ the continuity equation from which we obtain
\begin{equation}
\label{eq_defddft3}
 \frac{\partial \rho_{s\!/\!d}(\bvec{r},t)}{\partial t} = - \nabla\cdot \bvec{j}_{s\!/\!d}(\bvec{r},t) \, ,
\end{equation}
which completes the definition of the DDFT. The resulting theory with $\Gamma_0$ is obviously exact for non-interacting particles and it can be shown more generally to be exact provided that dynamic two-body correlations are identical to those in equilibrium \cite{ArEv04}. 


In practice, approximations have to be introduced already on the level of the equilibrium free energy functional $\mathcal{F}[\rho_s,\rho_d]$ and generally in a dynamical system two-body correlations differ from the respective equilibrium quantities. Therefore, DDFT provides an {\em approximation} to the evolution of the van Hove distribution functions in a system with realistic interactions.

The equilibrium density functional of the hard-sphere fluid is best described by FMT type expressions for the excess (over the ideal gas) free energy density $\Phi = \Phi(\{n_{\alpha}^{s\!/\!d}\})$, where $\{n_{\alpha}^{s\!/\!d}\}$ is a set of weighted densities which are obtained from a convolution of $\rho_{s\!/\!d}(\bvec{r},t)$ with given weight functions $\omega_{\alpha}$ \cite{Rf89}.
The full free energy functional reads
\begin{eqnarray}
  \mathcal{F}[\rho_s(\bvec{r},t),\rho_d(\bvec{r},t)] & = & \sum_{i=s, d} \int\! \upd\bvec{r} \,\rho_i(\bvec{r},t) (\ln \rho_i(\bvec{r},t) - 1)\nonumber \\ & & + \int\! \upd\bvec{r}\, \Phi(\{n_{\alpha}^{s\!/\!d}(\bvec{r},t)\})\nonumber \\ & &  +  \sum_{i=s, d}\int \!\upd\bvec{r} \,\rho_i(\bvec{r},t) V_i(\bvec{r},t) \, ,
\end{eqnarray}
where the first term on the r.h.s.\ corresponds to the free energy of the ideal gas with the thermal wavelength set to unity and the third term stems from the external potential $V_{s\!/\!d}(\bvec{r},t)$ acting on the respective species. Equilibrium density profiles for given chemical potentials $\mu_{s\!/\!d}$ are obtained via functional minimization of the grand potential functional $\Omega = \mathcal{F} - \sum_{i = s, d}\int \upd\bvec{r} \rho_i(\bvec{r})\mu_i$.

\begin{figure}[tbp]

  \includegraphics[width = 7.5cm]{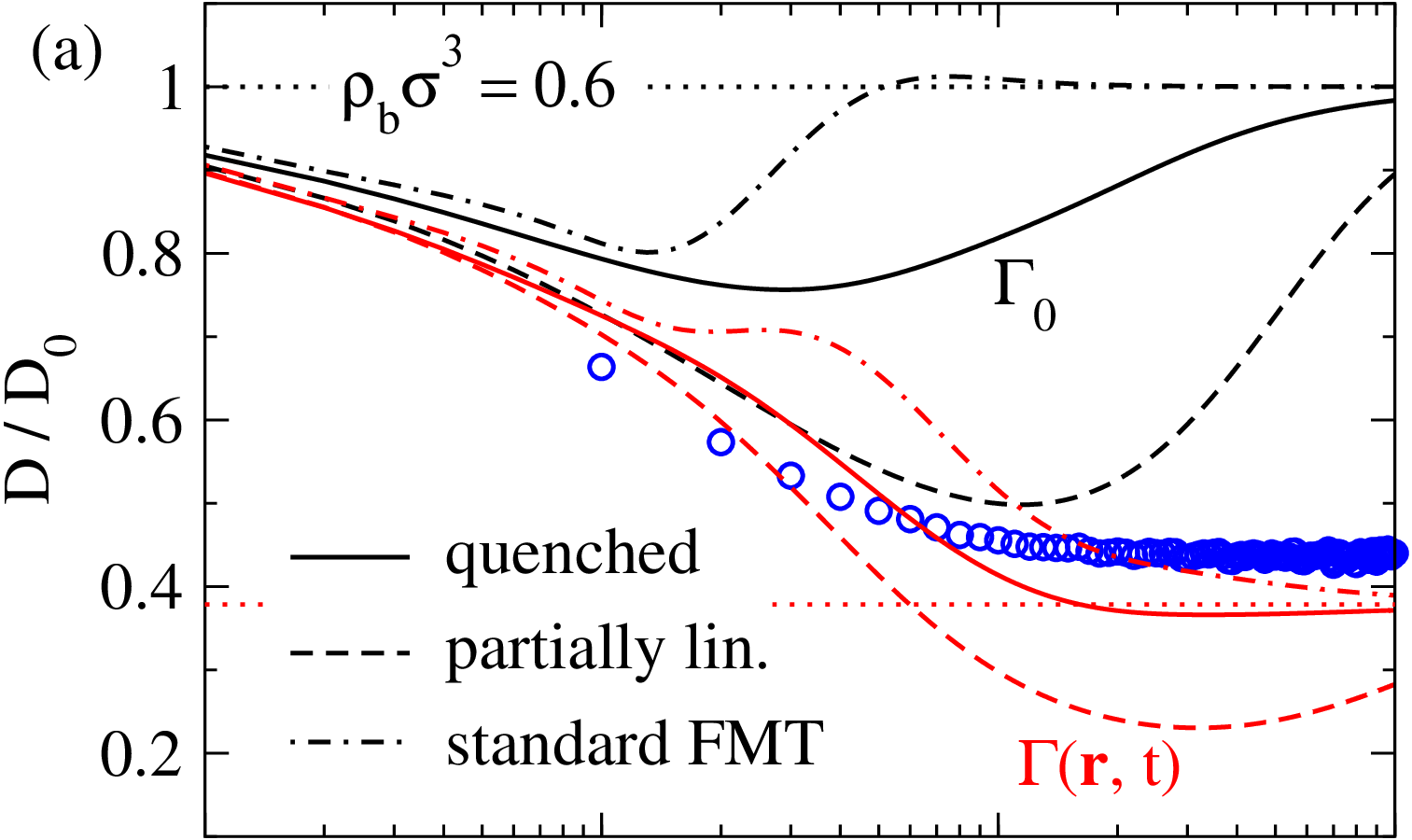} 

  \includegraphics[width = 7.5cm]{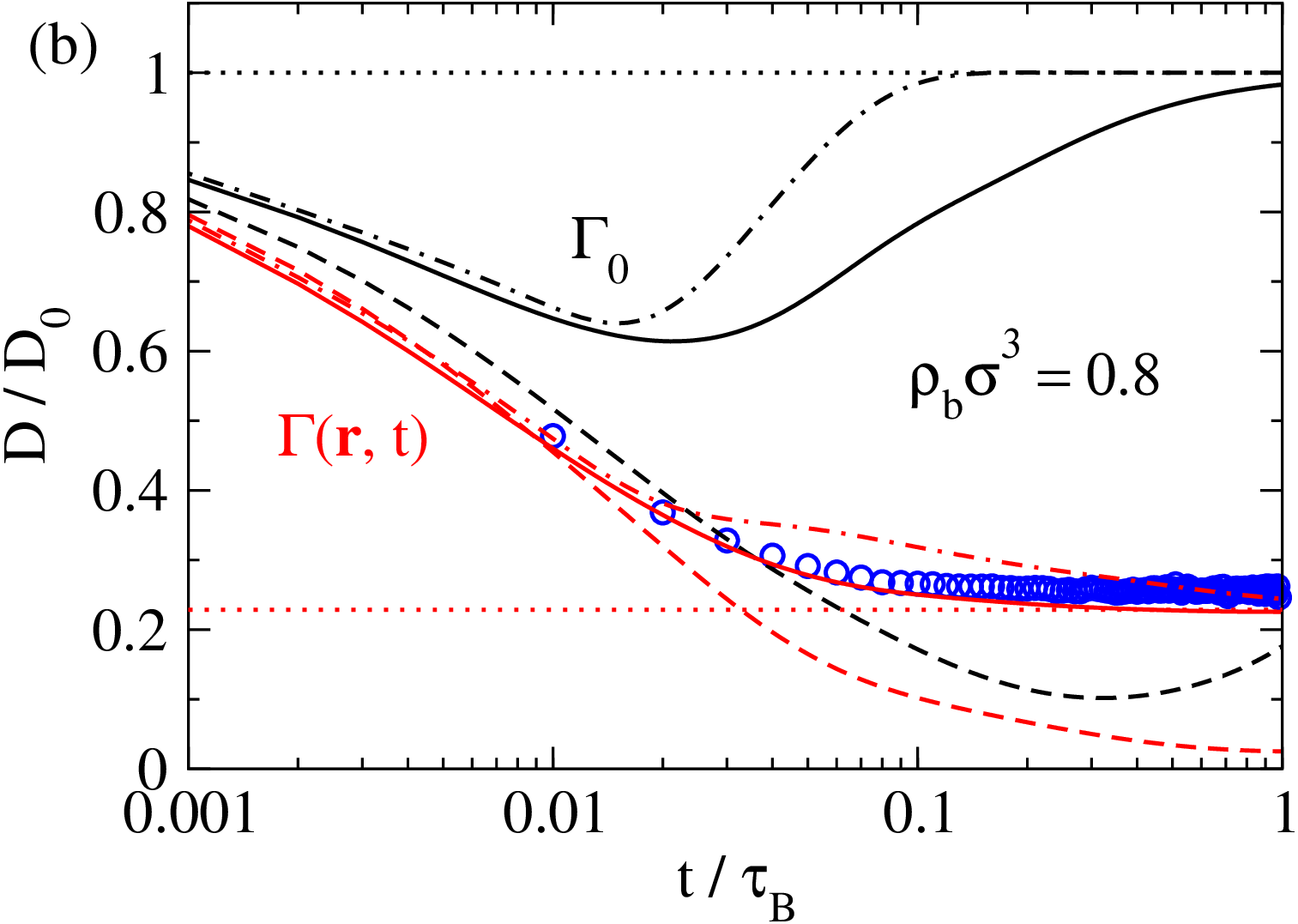} 

  \caption{Diffusion coefficient $D(t)$ relative to its value for the ideal gas $D_0$ calculated as the derivative of the MSD w.r.t.\ $t$. The MSD is obtained by numerically integrating $r^2\rho_s(r, t)$. Bulk densities are (a) $\rho_b\sigma^3=0.6$ and (b) $\rho_b\sigma^3=0.8$. The blue symbols are DMC data from Ref.~\onlinecite{Stopper15}. The quenched, partially linearized, and standard FMT free energy model were used in the DDFT, Eqs.~\eqref{eq_defddft1}-\eqref{eq_defddft3}, respectively; see text for details. The mobility in Eq.~\eqref{eq_defddft2} was that of the ideal gas (black curves) or that of Eq.~\eqref{eq_defGamma} (red curves). The dotted lines represent the values reached at long times.}\label{fig_DiffAll}
  
\end{figure}

The standard FMT type functional gives rise to interactions between and within the two components. However, in the present setting the self component contains only a single particle. Hence interactions within the self component have to be removed. In order to construct a functional in which the tagged particle does not interact with itself, we start with a quasi 0D cavity, i.e., a cavity which can hold at most one particle. The excess free energy of the system in the grand canonical ensemble is given by \cite{RoSchLoeTa97}
\begin{equation}
  \beta F_{\text{0D}}(\eta) = \eta + (1-\eta) \ln (1-\eta) \, ,
\end{equation}
where $0\le\eta\le1$ is the average occupation number of the cavity, i.e., the 0D packing fraction. This result has been instrumental in the construction of powerful free energy models for the inhomogeneous hard-sphere system, including the hard-sphere crystal \cite{Tar00}. The general procedure for the derivation of a FMT free energy functional based on a given 0D free energy is given in Ref.~\onlinecite{SchBrCP00}. The formalism includes binary mixtures, which is convenient in the context of the DDFT representation of the self and distinct parts of the van Hove distribution function $G(r,t)$.
Therefore, we introduce two packing fractions $\eta_s$ and $\eta_d$ in order to construct a suitable 0D free energy. In a first attempt we use
\begin{equation}
  F_{\text{0D}}^{\text{Rf}}(\eta_s,\eta_d) = F_{\text{0D}}(\eta_s+\eta_d) \, .
\end{equation}
The resulting functional corresponds to Rosenfeld's original FMT, or more precisely the subsequently developed version due to Tarazona \cite{Tar00}. This approach takes into account hard-sphere interactions within the distinct components, between the self and distinct components, but also {\em within} the self component. As a result, packing constraints are accurately captured by the approach, however, interactions within the self component are spurious considering that the latter consists of {\em exactly} one particle. In terms of the dynamics, this is reflected in an unrealistically fast broadening of the peak pertaining to the self component, resulting in a diffusivity which is too large compared to dynamic Monte Carlo (DMC) simulations (see dashed-dotted lines in Fig.~\ref{fig_DiffAll}). Note that throughout this work we use the White Bear II version of FMT \cite{HaRo06} with Tarazona's tensorial weighted densities \cite{Tar00}.

An alternative formulation, which removes interactions within the self component, consists in adopting the functional for colloid-polymer mixtures, which was put forward in Ref.~\onlinecite{SchBrCP00}. The interacting colloids are identified with the distinct component, while the non-interacting polymers represent the self component. The respective 0D free energy is obtained by linearizing $F_{\text{0D}}^{\text{Rf}}$ with respect to $\eta_s$. This free energy describes a 0D cavity which can hold either one distinct particle and no self particle, or no distinct particle and an arbitrary number of self particles, or no particles at all. Clearly, even for the average occupation number $\eta_s=1$ there is a significant statistical weight in the grand canonical ensemble for configurations where no self particle is present and hence occupation with a distinct particle is possible. As a result, the local packing constraint $\eta_s+\eta_d \le 1$ is violated. When employed within DDFT the resulting functional gives rise to what was termed the fully linearized functional in Ref.~\onlinecite{Stopper15}. It was shown that the distinct components loses its structure much too rapidly as it can easily fill in the space which in the physical system is always occupied by the self particle.

In order to find a compromise between the two scenarios, (i) the original FMT functional with a self peak that decays too rapidly, and (ii) the fully linearized functional with a distinct part that loses its structure too rapidly, the authors have recently suggested to use the so-called partial linearization route: the linearized functional is used to compute the one-body direct correlation function $c^{(1)}_s$ for the self component while $c^{(1)}_d$ of the distinct component is calculated using the original FMT functional. While this approach provides an accurate description of $G(r,t)$ as obtained in numerical simulations \cite{Stopper15}, it seems worthwhile to construct a single functional to be used within DDFT which (i) is free of interactions within the self component while (ii) providing an accurate representation of packing constraints.

A functional with the desired properties can be constructed by considering a 0D cavity with a quenched self particle. This corresponds to averaging the grand potential over configurations with and without a self particle being present. Hence the resulting grand potential $\Omega_q$ (quenched) is obtained as
\begin{equation}
  \Omega_q = \langle \Omega_{\nu} \rangle_{\nu=0,1} = \eta_s \Omega_0 + (1-\eta_s) \Omega_1 \, ,
\end{equation}
where $\Omega_0$ denotes the grand potential of a cavity which cannot be occupied by a distinct particle due to the presence of the self particle, while $\Omega_1$ corresponds to a cavity which can readily be occupied by a distinct particle. Note that the distinct component is coupled to a reservoir with a chemical potential $\mu_d$. Denoting $Z_1$ the canonical partition function of a single particle in the cavity, we have (up to an additive constant)
\begin{equation}
 \Omega_{\nu} = -\beta^{-1} \ln(1+\nu Z_1 e^{\beta \mu_d}) \, .
\end{equation}
From $\eta_d = -\frac{\partial \Omega_q}{\partial \mu_d}$ we obtain $Z_1 e^{\beta\mu_d} = \frac{\eta_d}{1-\eta_s-\eta_d}$. This result respects the packing constraint $\eta_s+\eta_d\le 1$. The free energy now follows from $F_q = \Omega_q + \mu_d \eta_d$. The excess free energy is obtained by subtracting the ideal gas contribution $\beta F^{\text{id}}=-\eta_d + \eta_d \ln(\eta_d/Z_1)$. The resulting excess free energy reads
\begin{eqnarray}
  \beta F_q^{\text{ex}}(\eta_s,\eta_d) & = & \eta_d + (1-\eta_s-\eta_d) \ln(1-\eta_s-\eta_d) \nonumber \\ & & - (1-\eta_s) \ln(1-\eta_s)\nonumber  \\
  & \equiv & \beta F_{\text{0D}}(\eta_s+\eta_d) - \beta F_{\text{0D}}(\eta_s) \, .
\end{eqnarray}
This result is very simple and has a number of appealing properties. In the limit of $\eta_d\to 0$, i.e., when no distinct particles are present, we have $F_q^{\text{ex}}=0$, hence interactions within the self component are removed. For $\eta_s\to 0$ we recover the standard 0D excess free energy $F_{\text{0D}}$. Moreover, owing to the term involving the logarithm of $1-\eta_s-\eta_d$ the local packing constraint is respected. Note that $F_q^{\text{ex}}$ does contain interactions within the self part provided that particles of the distinct component are present. These interactions can be viewed as mediated by the distinct component.

Following Ref.~\onlinecite{SchBrCP00} an FMT excess free energy functional $\mathcal{F}_q^{\text{ex}}$ can be constructed from $F_q^{\text{ex}}$. Due to the linearity of the operations that are applied, it follows that
\begin{equation}
\label{eq_Fqfunc}
   \mathcal{F}_q^{\text{ex}}[\rho_s(\bvec{r}),\rho_d(\bvec{r})] =  \mathcal{F}^{\text{ex}}[\rho_s(\bvec{r})+\rho_d(\bvec{r})]
-\mathcal{F}^{\text{ex}}[\rho_s(\bvec{r})] \, ,
\end{equation}
where $\mathcal{F}^{\text{ex}}$ is the usual FMT excess free energy. Obviously, Eq.~\eqref{eq_Fqfunc} is a statement that can directly be generalized to arbitrary (non-FMT) functionals. Reinhardt and Brader use a similar approach in order to eliminate self-interactions of hard rods in $d=1$, however they do not provide a derivation based on physical principles \cite{ReiBr12}.

In Fig.~\ref{fig_DiffAll} we plot the diffusion coefficient $D(t)$ as obtained from the new functional $\mathcal{F}_q^{\text{ex}}$, the standard FMT functional, and the partial linearization route of Ref.~\onlinecite{Stopper15}. While for short times free diffusion is correctly described, all three functionals yield the diffusion coefficient of the ideal gas for long times. In order to fix this shortcoming, we use a result obtained by Leegwater and Szamel \cite{LeSz92} for the long-time mobility $\Gamma(\eta)$ of the hard-sphere fluid of packing fraction $\eta$. Using ideas of Enskog kinetic theory, they obtain the expression
\begin{equation}
\label{eq_genGamma}
  \Gamma(\eta) = \frac{\Gamma_0}{1 + 2 \eta g(\sigma^+)} \, ,
\end{equation}
where $\Gamma_0$ is the mobility of the spheres in the dilute limit and $g(\sigma^+)$ is the radial distribution function at contact, itself a function of $\eta$. The latter can be calculated from the FMT type free energy within morphometric thermodynamics \cite{KoeRoMe04}. 
Using the White Bear II version of FMT \cite{HaRo06} we obtain an analytical formula yielding a precise value for $g(\sigma^+)$ at a given packing fraction $\eta$ to be used in Eq.~\eqref{eq_genGamma}. It remains to specify at which packing fraction Eq.~\eqref{eq_genGamma} has to be evaluated given an arbitrary inhomogeneous configuration of the fluid at a certain time $t$. In the spirit of Ref.~\onlinecite{RoyallEA07}, we decide for the most natural choice within FMT, which is to define a local packing fraction based on the weighted density $n_3$, which integrates the fluid density over the volume of a sphere. Hence in the bulk fluid $n_3=\eta$ while in the inhomogeneous fluid we have $0\le n_3 \le 1$. More specifically we set
\begin{equation}
\label{eq_defGamma}
  \Gamma_s(\bvec{r}, t) \doteq \Gamma(n_3^{d}(\bvec{r}, t))\, , \,\, \Gamma_d(\bvec{r}, t) \doteq \Gamma(n_3^{s}(\bvec{r}, t)+n_3^{d}(\bvec{r}, t)) \, .
\end{equation}
These definitions take into account that there are no interactions within the self-component, consequently the respective mobility depends only on $n_3^{d}(\bvec{r})$. Obviously, in the long-time limit where $n_3^{s} \to 0$ and $n_3^{d} \to \eta$ the correct values for the mobility are obtained. More interestingly, the definition insures that for short times the self particle can diffuse freely. This is a result of $n_3^{d}$ vanishing at the initial location of the self-particle for short times. Hence $\Gamma_s=\Gamma_0$ in the vicinity of $\bvec{r} = 0$ in the short time limit.


In Fig.~\ref{fig_DiffAll} we show the diffusivity resulting from the use of the mobility from Eq.~\eqref{eq_defGamma} (red curves). As expected, the asymptotically reached mobility for large $t$ is significantly improved w.r.t.\ the DDFT using $\Gamma_0$ which yields the ideal gas diffusivity in the long term. However, for $\rho_b \sigma^3 = 0.6$ the long-time diffusivity born out by the modified DDFT is somewhat too small, which reflects the inaccuracy of the simple formula by Leegwater and Szamel, Eq.~\eqref{eq_genGamma}. Importantly, the power of the functional $\mathcal{F}_q^{\text{ex}}$ defined in Eq.~\eqref{eq_Fqfunc} is now clearly visible. The functional from the partial linearization route leads to diffusion that is much too slow while the functional which contains interactions within the self component yields a diffusion that is too fast compared to simulations. Only $\mathcal{F}_q^{\text{ex}}$, in which interactions within the self component have been removed using the ``quench'' route, gives a diffusivity that compares very well with simulations. Regarding the satisfactory performance of the partial linearization route, in particular at short and intermediate times \cite{Stopper15}, we can conclude that this is the result of a fortunate compensation of two opposite effects: partial linearization slows down the dynamics significantly, while using the mobility of the ideal gas speeds it up sufficiently to create reasonable agreement with simulations except in the long-time limit. On the other hand, it is clearly seen in Fig.~\ref{fig_DiffAll} that partial linearization {\em with} the correct mobility (dashed red lines) provides a poor account of simulations. The corresponding curves display an unphysical minimum where the diffusivity becomes unrealistically low before asymptoting to the Leegwater-Szamel value. This is particularly pronounced for $\rho_b \sigma^3 = 0.8$ with a minimum value of $D \approx 0.025 D_0$ at $t \approx 1.12 \tau_B$.

\begin{figure}[tbp]
    
  \includegraphics[width = 7.5cm]{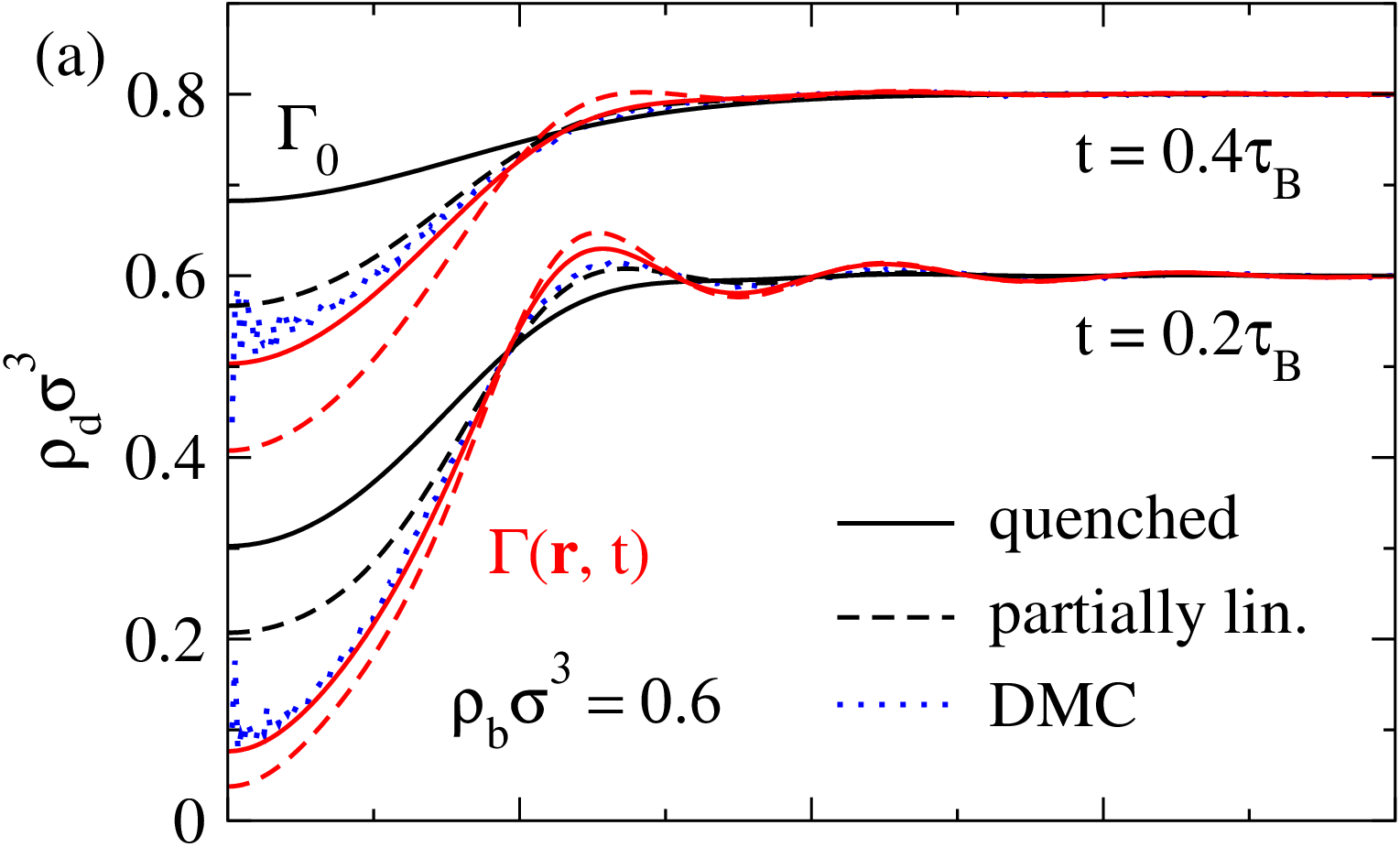} 

  \includegraphics[width = 7.6cm]{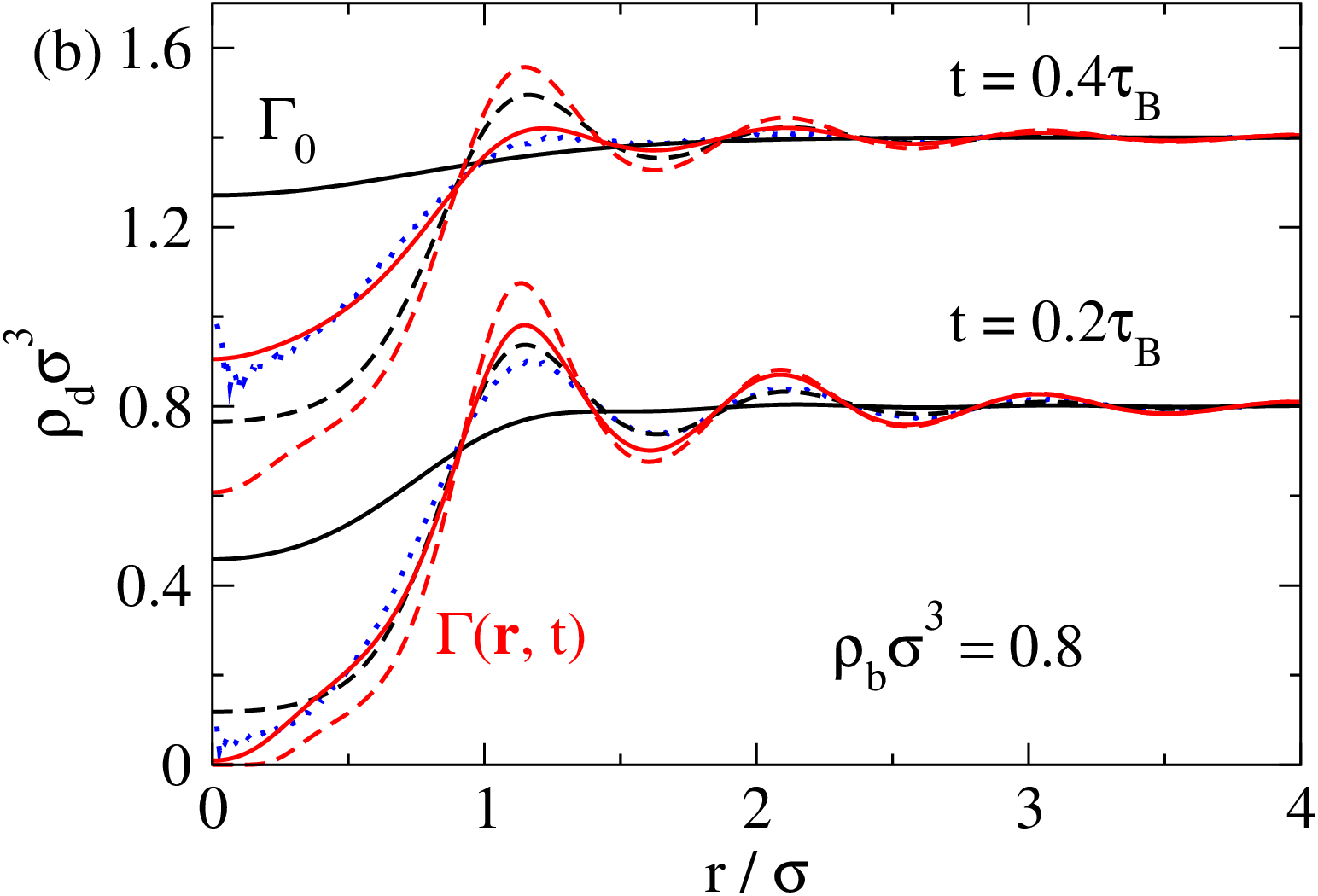} 

  \caption{Distinct part of the van Hove distribution function $G_d(r, t) \equiv \rho_d(r,t)$ for bulk densities (a) $\rho_b\sigma^3=0.6$ and (b) $\rho_b\sigma^3=0.8$ at times $t=0.2\tau_B$ and $t=0.4\tau_B$ as obtained from DMC simulations and different DDFT approaches. The curves for $t=0.4\tau_B$ have been shifted for clarity. The partial linearization approach of Ref.~\onlinecite{Stopper15} is compared to the present ``quenched'' DDFT approach. The particle mobility is either that of the ideal gas (black curves) or the modified expression from Eq.~\eqref{eq_defGamma} (red curves).}\label{fig_profilesAll}
  
\end{figure}

Finally, we show results for the distinct component $\rho_d(\bvec{r},t)$ in Fig.~\ref{fig_profilesAll}. As expected from the diffusivity results the relaxation of $\rho_d(\bvec{r},t)$ to the homogeneous bulk density is generally too fast for the DDFT using the ideal gas mobility $\Gamma_0$ (black curves). Even at $\rho_b \sigma^3 = 0.8$, Fig.~\ref{fig_profilesAll} (b), where for the partial linearization balances the speed up at the times shown here (while still yielding dynamics that are too fast in the long-time limit) the overall agreement is clearly poorer than what is obtained with the mobility modified according to Eq.~\eqref{eq_defGamma} and the functional $\mathcal{F}_q^{\text{ex}}$ from Eq.~\eqref{eq_Fqfunc} (solid red curves). The latter agrees almost perfectly with simulations, especially near the origin where the tagged particle is located at $t=0$. The only noticeable difference is that the oscillations in $\rho_d(\bvec{r},t)$ given by the DDFT are too pronounced at $t=0.2\tau_B$. This slight deviation, however, vanishes at larger times (see curves for $t=0.4\tau_B$). 

Even at $\rho_b \sigma^3 = 1$ (not shown here) the van Hove function is well described by the present DDFT as we infer from comparison with MD simulation results from Ref.~\onlinecite{HopEA10}. Again the oscillations of the distinct part decay somewhat too slowly at short times but by $t=0.1\tau_B$ the agreement is already very good.

Interestingly, unlike previous DDFT formulations \cite{HopEA10,Stopper15}, the present DDFT does not predict dynamic arrest even for systems with packing fractions exceeding that of the random close packing ($\sim 64\%$). This strengthens the previous conjecture \cite{Stopper15} that dynamic arrest observed with standard DDFT in the past is merely an artifact of poor free energy models \cite{HopEA10}. We therefore conclude that a dynamic phenomenon such as the glass transition most likely cannot be described within DDFT as long as the mobility in Eq.~\eqref{eq_defddft2} does not vanish at a given critical density. In order to study the glass transition without empirical input of a suitable mobility, novel approaches such as for instance power functional theory \cite{SchBr13} seem to be necessary.
The present findings encourage work along those lines since standard DDFT has been shown here to be insufficient to generate dynamic arrest.


In this Communication, we have presented a DDFT for Brownian motion of hard-sphere like colloids which makes use of the FMT free energy functional for the hard-sphere fluid. Two crucial modifications have to be applied in order to obtain good agreement with simulations: interactions within the self part of the van Hove function have to be removed in an approach using a 0D cavity with a quenched self component, {\em and} a particle mobility which is a function of space and time via a locally averaged packing fraction must replace the mobility of the ideal gas. The present model gives an account of Brownian motion for a system with non-trivial interactions, that is {\em quantitatively} accurate even at a packing fraction of $40\%$ and semi-quantitatively accurate for packing fractions of $50\%$. The route is now open for extensions of the model toward more complex interactions (van der Waals, hydrodynamic) and studies involving confining (time-dependent) potentials which are present in many experiments with colloidal suspensions.

\end{document}